\documentclass[epj]{webofc}
\usepackage[varg]{txfonts}   
\woctitle{MESON2018 - the 15$^\textrm{th}$ International Workshop on Meson Physics}
\begin{document}
\selectlanguage{english}
\title{Heavy mesons in the Quark Model}

\author{D.R. Entem\inst{1}\fnsep\thanks{\email{entem@usal.es}} \and
        P.G. Ortega\inst{1} \and
        J. Segovia\inst{2}  \and
	F. Fern\'andez\inst{1}
}

\institute{Grupo de F\'isica Nuclear and Instituto Universitario de F\'isica
Fundamental y Matem\'aticas (IUFFyM), Universidad de Salamanca, E-37008 Salamanca, Spain
\and
           Grup de F\'isica Te\`orica, Dept. F\'isica and IFAE-BIST, Universitat Aut\`onoma de Barcelona, E-08193 Bellaterra (Barcelona), Spain
          }

\abstract{
Since the discovery of the $J/\psi$, the quark model was very successful in describing the spectrum and properties of heavy mesons including only $q\bar q$ components. However since 2003, with the discovery of the $X(3872)$, many states that can not be accommodated on the naive quark model have been discovered, and they made unavoidable to include higher Fock components on the heavy meson states. We will give an overview of the success of the quark model for heavy mesons and point some of the states that are likely to be more complicated structures such as meson-meson molecules.
}
\maketitle
\section{Introduction}

In November 1974 the Brookhaven National Laboratory~\cite{prl33.1404.1974} announced the discovery of a new particle, called $J$. 
Almost at the same time the existence of another new particle, called $\psi$,  was reported by the Stanford Linear Accelerator~\cite{prl33.1406.1974}. 
Both particles shared similar properties and were interpreted as a $q\bar q$ state made of a new quark, the charm quark, previously predicted by the 
GIM mechanism~\cite{prd2.1285.1970} which explained the suppression of flavor changing kaon weak decays. These discoveries started a fast 
development of particle physics that is known as the November Revolution.

Very soon after, the $\Upsilon(1S)$ was discovery at Fermilab~\cite{prl39.252.1977} at higher energies confirming the existence of a fourth quark, 
the bottom quark. Until 1980, 11 states were collected by the Particle Data Group in these energy ranges~\cite{rmp52.s1.1980} (see Table~\ref{tab1}).

\begin{table}[!htb]
\centering
\caption{States included in the Particle Data Group in 1980~\cite{rmp52.s1.1980}.}
\label{tab1}       
\begin{tabular}{lll}
\hline
      & Mass (MeV) & $J^P$  \\\hline
$J/\psi(3100)$ & $3097 \pm 1$ & $1^-$ \\
$\chi(3415)$ & $3414 \pm 4$ & $0^+$ \\
$p_c$ or $\chi(3510)$ & $3507 \pm 4$ &  \\
$\chi(3550)$ & $3551 \pm 5$ &  \\
$\psi(3685)$ & $3685 \pm 1$ & $1^-$ \\
$\psi(3770)$ & $3768\pm3$ & $1^-$ \\
$\psi(4030)$ & $4030 \pm 6$ & $1^-$ \\
$\psi(4160)$ & $4159 \pm 20$ & $1^-$ \\
$\psi(4415)$ & $4415 \pm 6$ & $1^-$ \\
$\Upsilon(9460)$ & $9458 \pm 6$ & $1^-$ \\
$\Upsilon(10020)$ & $10016 \pm 14$ & $1^-$ \\ \hline
\end{tabular}
\end{table}

In 1978 the Cornell model~\cite{prd17.3090.1978} was developed and it already took into account the most important features of the quark-antiquark 
interaction in the heavy sector. The basic assumptions of the model were that the interactions were governed by $SU(3)$ color gauge symmetry 
with flavor only broken by the quark masses. It includes a coulomb term, induced by a one-gluon exchange interaction, and a phenomenological 
confining interaction, that was taken to be linear. The interactions were flavor independent and spin independent, implementing the well 
known nowadays Heavy Flavor Symmetry and Heavy Quark Spin Symmetry.

The original naive quark model from Cornell was very successful in explaining the charmonium and bottomonium spectrum~\cite{prd21.203.1980}. 
The model parameters were fitted to the 11 states mentioned before. The model gave predictions mainly for the bottomonium spectrum and
comparing with the data of the PDG 2003~\cite{prd66.010001.2002}, where 15 states were added, all the predictions were on the correct energy range. 
In order to give a better description of the spectrum more elaborate models were developed being some of the most representative the model 
of Godfrey and Isgur~\cite{prd32.189.1985} and the model of Ebert {\it et al.}~\cite{epjc71.1825.2011}.

The situation changed completely in 2003 with the discovery of the $X(3872)$ by the Belle Collaboration~\cite{prl91.262001.2003} and 
very soon after confirmed by the CDF~\cite{prl93.072001.2004}, D0~\cite{prl93.162002.2004} and BaBar~\cite{prd71.071103.2005} Collaborations. 
This state could decay into $J/\psi \rho$, an isospin 1 channel, which ruled out completely a pure $c\bar c$ interpretation. 
Its closeness with the $D^0\bar D^{*0}$ threshold suggested a molecular interpretation and clearly showed that the naive quark model 
was not enough to explain this state.

The number of states on the charmonium and bottomonium region has increase since 2003 and in the PDG 2017~\cite{prd98.030001.2018} a 
total of 57 states are quoted, some of them with a clear non-$q\bar q$ nature.

\section{Two meson dynamics and the Chiral Quark Model}

In order to understand the $X(3872)$ the two meson dynamics is necessary. The first thing to notice is that
quark models for heavy mesons only include color interactions. The interaction generated between mesons for
such interactions between quarks cancels since mesons are colorless objects. However in the case of open charm meson-antimeson 
states the interaction between light quarks is also relevant.

In the light quark sector another important feature of QCD arises. The QCD Lagrangian is Chiral symmetric for massless quarks.
However Chiral symmetry is not realized in the light meson sector, and the small masses of light quarks can not explain
the breaking observed. This breaking is understood due to the spontaneously symmetry breaking of the QCD vacuum which generates
the appearance of Goldstone bosons. For exact symmetry the Goldstone bosons are massless, however in the case of two flavor QCD
they are light bosons which we can clearly identify with pions. The important point here is that now light quarks can
interact through the exchange of Goldstone bosons, and these interactions are colorless, so they don't cancel between colorless
objects. This is a way to build the two meson dynamics from the quark interaction.

With these ideas the Chiral Quark Model was developed~\cite{jpg19.2013.1993} and extended to $SU(3)$~\cite{jpg31.481.2005}.
It has extensively used in the literature to study 
the $NN$ interaction~\cite{prc62.034002.2000},
the baryon spectrum~\cite{plb367.35.1996} and
the meson spectrum~\cite{jpg31.481.2005,prd78.114033.2008,prd93.074027.2016}. 

\begin{figure}[t]
\centering
\includegraphics[width=10.2cm,clip]{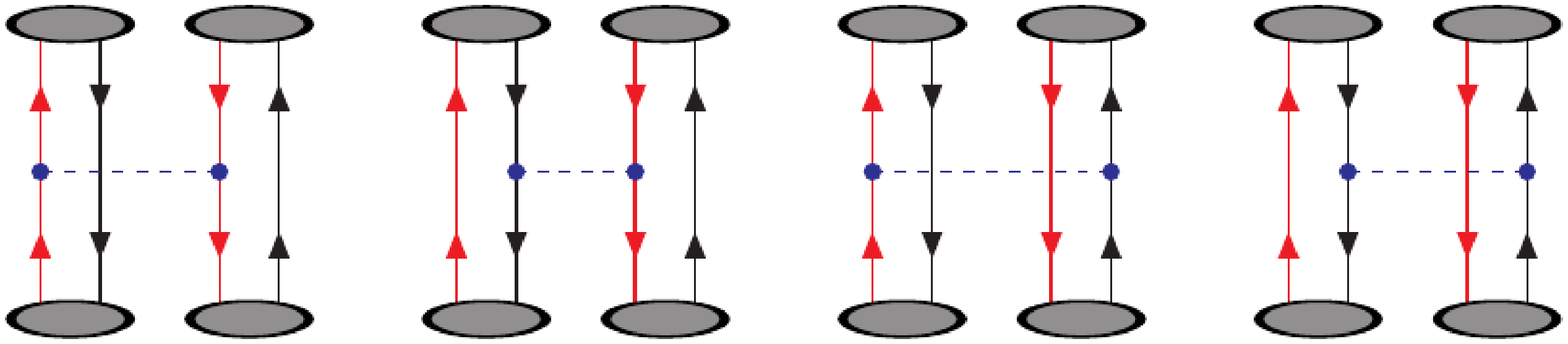}
\includegraphics[width=10cm,clip]{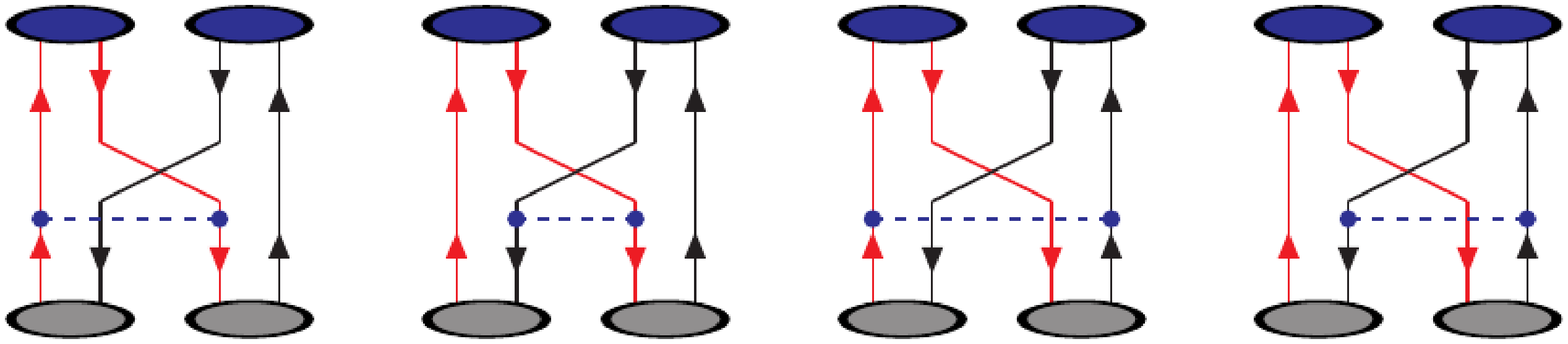}
\caption{Diagrams that contribute to the interaction between mesons and rearrangement processes.}
\label{fig1}       
\end{figure}

With the interaction between quarks we use the Resonanting Group Method (RGM)~\cite{pr52.1083.1937} to obtain the interaction between mesons
(diagrams on the first row of Fig.~\ref{fig1})
or transitions from a two meson channel to another
(diagrams on the second row of Fig.~\ref{fig1}), which are the so called rearrangement processes.

However the dynamics between two meson states is also governed by the coupling with one meson states. This is clear since one meson states 
decays strongly into two meson states through OZI allowed decays, and it was already taken into account in the original Cornell Model~\cite{prd17.3090.1978}.
A simple an efficient way to account for such processes is to used a phenomenological $^3P_0$ model~\cite{npb10.521.1969}. It has been extensively used
in different systems and it only depends on a parameter that can run with the scale~\cite{plb715.322.2012}.

With all this in mind one can start from a wave function
\begin{equation}
	|\Psi\rangle = \sum_\alpha c_\alpha |\psi\rangle + \sum_\beta \chi_\beta(P) |\phi_{M_1}\phi_{M_2}\beta \rangle,
\end{equation}
where the first term takes into account the relevant $c\bar c$ bare states and the second incorporates the two meson components.

One ends up with a Schr\"odinger type equation
\begin{equation}
\sum_\beta \int \left( H_{\beta' \beta}^{M_1M_2}(P',P) +
V_{\beta'\beta}^{eff}(P',P)\right)\chi_\beta(P) P^2 dP=E\chi_{\beta'}(P')
\end{equation}
with
\begin{equation}
V_{\beta'\beta}^{eff}(P',P) = \sum_\alpha \frac{h_{\beta'\alpha}(P')h_{\alpha\beta}(P)}{E-M_\alpha}
\end{equation}
an effective potential between mesons due to the coupling with bare $q\bar q$ states, being $h_{\alpha\beta}(P)$ the vertex function given
by the $^3P_0$ model. Here it is important to notice that the states that give a higher contribution are those close to the threshold,
being attractive when the state is above and repulsive when the state is below.

\section{Threshold effects in the heavy meson spectrum}

There are many states that can be accommodated in the naive quark model or given basically by a two meson molecule. However this is not always the case and
an example is the already mentioned $X(3872)$. As explained before this state has properties that rules out a naive quark model interpretation, however
it could be a two meson molecule. Its nature has important consequences since we have in hand important symmetries as Heavy Quark Spin Symmetry and
Heavy Flavor Symmetry. This has been studied in the pure molecular picture in Refs.~\cite{prd86.056004.2012,prd88.054007.2013}. The unavoidable consequence
is that the existence of the $X(3872)$ in the $1^{++}$ sector implies the existence of a partner in the $2^{++}$ channel in the charmonium sector, and the
existence of their analogs in the bottomonium sector. None of this states have been found and in particular the bottom analog of the $X(3872)$
has been searched by
the CMS~\cite{plb727.57.2013}, the ATLAS~\cite{plb740.199.2015} and the Belle~\cite{prl113.142001.2014} Collaborations and no state was found.

This fact suggests that the state should have a different nature. In the energy region of the $X(3872)$ most quark models predict states with a higher
mass around 3910. However the closeness of the $DD^*$ threshold implies that there should be a mixing between naive quark model states and two meson
components. It has been shown in Ref.~\cite{prd81.054023.2010} that, in the Chiral Quark Model, a state near the $D^0D^{*0}$ threshold appears when
the coupling with the naive $\chi_{c1}(2P)$ state is considered, since this state is close enough to give a sizeable attraction. 
A similar calculation at baryon
level has been carried out in Refs.~\cite{epjc76.576.2016,PTEP2013.093D01.2013}.

In this picture the consequences of HQSS and HFS are different. The model has these two symmetries~\cite{aip1735.060006.2016} in the heavy sector. However
the mass difference between the $D$ and $D^*$ mesons, which is around 100 MeV, plays an important role since the relative position between the naive quark 
model states and the relevant thresholds changes from one channel to the other. This energy region has been analyzed in detailed in Ref.~\cite{plb778.1.2018}.
There it was demonstrated that the $2^{++}$ analog does not appear within the model. The naive quark model states get dressed but only an additional state
is found in the $1^{++}$ channel (the X(3872)) an one in the $0^{++}$ channel. 

The $2^{++}$ state is interesting because its decay properties suggest that the states $X(3915)$ and $X(3930)$ could be the same state with these quantum 
numbers. This situation agrees with Ref.~\cite{prd83.114015.2011}.

In the $0^{++}$ channel two states appear. In this channel there were originally two experimentally measured states,
the $Y(3940)$ and the $X(3915)$, which later on were seen as the same
by the community. It was even relabeled by the PDG as the $\chi_{c0}(2P)$ although this assignment has been abandoned. If the $X(3915)$ is the $2^{++}$
state then one of the $0^{++}$ could be the $Y(3940)$ and the other a recently measured resonance called $X(3860)$.

However, as mentioned before, threshold effects are not always essential. Recently the LHCb Collaboration measured four new 
resonances~\cite{prl118.022003.2017,prd95.012002.2017}, namely, the 
$X(4140)$, $X(4274)$, $X(4500)$ and $X(4700)$, with quantum numbers $1^{++}$ for the first two states and $0^{++}$ for the last two states. As these
states are well above threshold one would expect that the coupling with two meson states were very important. In Ref.~\cite{prd94.114018.2016} 
these states have been
analyzed. Results indicate that the $X(4274)$ could be basically the $\chi_{c1}(3P)$ state with some dressing from nearby thresholds. The same
situation is found for the $X(4500)$ that could be the $\chi_{c0}(4P)$ and the $X(4700)$ the $\chi_{c0}(5P)$.
For the $X(4140)$, that could be a threshold cusp, no candidate has been found.

\section{Summary}

The naive quark model has been very successful at explaining the spectrum of heavy mesons in the charmonium and bottomonium regions. However with the
discovery of the $X(3872)$ in 2003 it became clear that more complex structures than quark-antiquark states should be present. Although the naive
quark model represents a good guidance to understand the spectrum, the mixing with nearby two meson thresholds have to be taken into account.

Some times the mixing only represents a dressing of naive quark model states, without a big change of their properties. However, in some cases, the
mixing implies the existence of additional states with respect to the ones predicted by the naive quark model.

On the other hand the picture for two meson molecules could be quite different when mixing with one meson states are taken into account. In particular
expected symmetries as HQSS and HFS sometimes are not realized and results differ drastically from the expectations due to these symmetries.

\begin{acknowledgement}
This work has been partially funded by Ministerio de Econom\'ia, Industria y
Competitividad under Contracts No. FPA2016-77177-C2-2-P, FPA2014-55613-P, FPA2017-86989-P and SEV-2016-0588 and by Junta de Castilla y Le\'on and ERDF under Contract No. SA041U16.
P.G.O. acknowledges the financial support from Spanish MINECO's Juan de la Cierva-Incorporaci\'on programme, Grant Agreement No. IJCI-2016-28525. J.S. acknowledges the financial support from the European Union's Horizon 2020 research and innovation programme under the Marie Sk\l{}odowska-Curie Grant Agreement No. 665919, and from Spanish MINECO's Juan de la Cierva-Incorporaci\'on programme, Grant Agreement No. IJCI-2016-30028.
\end{acknowledgement}
%
%
%

\end{document}